\documentclass[12pt,preprint]{emulateapj}

\usepackage{color,txfonts}
\usepackage{graphicx}

\def\apj{{ApJ}}
\def\apjl{{ApJL}}

\def\be{\begin{equation}}
\def\ee{\end{equation}}
\def\bea{\begin{eqnarray}}
\def\eea{\end{eqnarray}}

\begin{document}

\title{Is the line-like optical afterglow SED of GRB 050709 due to a flare?}

\author{Cong Liu\altaffilmark{1,2}, Zhi-Ping Jin\altaffilmark{1} and Da-Ming Wei\altaffilmark{1}}

\altaffiltext{1}{Key Laboratory of Dark Matter and Space Astronomy, Purple Mountain Observatory, Chinese Academy of Sciences, Nanjing 210008, China }
\altaffiltext{2}{University of Chinese Academy of Sciences, Yuquan Road 19, Beijing, 100049, China}

\email{jin@pmo.ac.cn (ZPJ)}

\begin{abstract}
Recently Jin et al. (2016) reanalyzed the optical observation data of GRB 050709 and reported a line-like spectral energy distribution (SED) component observed by the Very Large Telescope at $t\sim 2.5$ days after the trigger of the burst, which had been interpreted as a broadened line signal arising from a macronova dominated by iron group. In this work we show that an optical flare origin of such a peculiar optical SED is still possible. Interestingly, even in such a model, an ``unusual" origin of the late-time long-lasting Hubble Space Telescope F814W-band emission is still needed and a macronova/kilonova is the natural interpretation.
\end{abstract}


\section{Introduction}

Gamma-ray bursts (GRBs) are short flashes of $\gamma$-rays from the outer space. In general GRBs can be divided into two distinct sub-groups based on the duration distribution of the prompt soft gamma-ray emission \citep{1993ApJ...413L.101K}. One group, with a typical duration of $\sim 20~{\rm s}$, has been named as ``long GRBs". The other group, with a much shorter duration $<2$ s, has been called the ``short GRBs" (sGRBs). Except a few outliers that have been called the long-short or hybrid GRBs, the nearby long GRBs were found to be associated by bright supernovae \citep{Kumar2015} and hence from the collapse of massive stars. In contrast, no bright supernova emission has been detected in the afterglow of sGRBs and hybrid GRBs. Such a fact has been taken as one of the most compelling evidence for the compact binary merger origin of these events \citep{Berger2014}.

Though no luminous supernovae in the afterglow of sGRBs and hybrid GRBs have been detected so far, the searches for much weaker and softer near-infrared/optical transients powered by the radioactive decay of
$r-$process material synthesized in ejecta launched during the mergers, i.e., the so-called Li-Paczynski macronovae/kilonovae \citep{Li1998,Kulkarni2005,Metzger2010}, have attracted wider and wider attention. In comparing with the remarkable progresses made on numerical simulation in these few years \citep[e.g.,][]{Roberts2011,Korobkin12,Barnes2013,Kasen2013,Tanaka2013,Piran2013,Grossman2013,Lippuner2015}, the macronova/kilonova sample increases rather slowly. In 2013 the first macronova/kilonova candidate, mainly based on one single Hubble Space Telescope (HST) F160W-band detection, was identified in sGRB 130603B \citep{Tanvir2013,Berger2013}. In 2015, the joint re-analysis of the Very Large Telescope (VLT) and HST afterglow data of the hybrid GRB 060614, a burst famous for sharing some characters of both long and short GRBs, revealed strong evidence for the presence of a macronova/kilonova component \citep{Yang2015}. Further examination of the afterglow allowed some authors to derive a tentative macronova light curve \citep{Jin2015}. Very recently, \citet{Jin2016} re-analyzed the VLT and HST afterglow data of sGRB 050709. Together with the Danish 1.5m telescope optical afterglow data \citep{Hjorth2005}, \citet{Jin2016} showed that the optical signal, interpreted as the afterglow, is actually dominated by the macronova/kilonova component at $t\gtrsim 2.5$ days and such a component is rather similar to that identified in hybrid GRB 060614.
Note that sGRB 050709 was the first short event with an identified optical counterpart, the successful identification of a macronova/kilonova emission component in the optical data implies that macronova/kilonova is common in merger-originated (either double neutron star merger or neutron star and black hole merger) events. The statistical study of the nearby short and hybrid GRBs with sufficient optical afterglow observations to hunt for macronova/kilonova provides indeed a strong support to such a possibility \citep{Jin2016}. Though these progresses are intriguing and encouraging, the ``peculiar" VLT $I/R/V$ afterglow emission measured at $t\sim 2.5$ days is still to be better understood. The peculiarity is that the R-band flux is significantly larger than the I-band and V-band fluxes, which is unexpected in the afterglow emission. \citet{Jin2016} hypothesized that  such a signal is due to a wind-macronova since a strong line feature can be produced by a macronova dominated by iron group \citep{Kasen2013}. If correct, it reveals the nature of the composition of the sub-relativistic outflow. However, we noticed that the VLT $I/R/V$ measurements were carried out individually. The different measure times actually left more freedom for the theoretical interpretation, which is the focus of this work.

This work is organized as the following. In Section 2 we show that the result of the VLT $I/R/V$ measurements at $t\sim 2.5$ days is indeed inconsistent with the regular afterglow model. We discuss in Section 3 an optical flare model to interpret the peculiar VLT optical spectral energy distribution (SED)  and in Section 4 we discuss whether  in this case a macronova component is still needed or not.  We summarize the results with some discussions in Section 5.

\section{The line-like VLT $I/R/V$ SED}
The optical counterpart was first detected by Danish 1.5m telescope, a series of observations were made but only the first two epoches of $R$ band observations at one day and two days after the burst detected the counterpart\citep{Hjorth2005}. Notable that \citet{Covino2006} reported the VLT detection of sGRB 050709 counterpart in $R$ and $V$ bands and an upper limit in $I$ band. They are a sequence of observation taken in about half an hour, first six 100-seconds exposures in $I$-band, then six 60-seconds exposures in $V$-band, and finally one 20-seconds and five 60-seconds exposures in $R$-band. Thanks to the more suitable choice of  now available  reference frame of the $I$-band data analysis,
the very-recent re-analysis by \citet{Jin2016} found a significant $I$-band detection. Later, three more epoches of observations were conducted, however no significant signed of optical counterpart were detected. The detection of the afterglow in $I/R/V$ bands in a short time interval by VLT on July 12, 2005 can be directly used to build the SED without any correction, as shown in Fig.1. In the regular forward shock model, the long-lasting afterglow emission is attributed to the synchrotron radiation of the electrons accelerated in the shock front and the spectrum is well understood \citep{Sari1998}. Usually the optical afterglow spectrum is power-law like. However when we try to fit the VLT SED by a power-law, the best fit yields a reduced $\chi^2$ of $\sim 12.4$ (see the yellow dashed line in Fig.1), too large to be acceptable. A thermal spectral fit yields a reduced $\chi^2$ of $\sim 10$, which is not acceptable either (see the blue dashed line in Fig.1).
Such facts strongly suggest that the VLT optical emission is irregular and specific model is crucially needed \cite[see also][]{Jin2016}, which is the focus of the rest of this work. HST also took three epoches of F814-band observations within a month after the burst, all clearly detected the counterpart. Later, two more epoches of observation were made, but the counterpart has faded away\citep{Fox2005}.

\begin{figure}
\label{fig:SED}
\begin{center}
\includegraphics[clip, width=3.3 in]{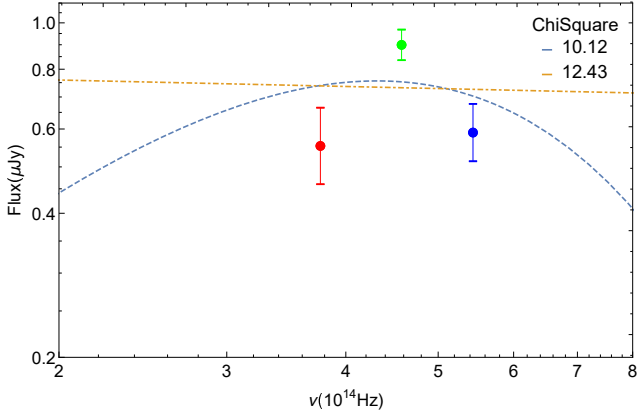}
\caption{The SED of the afterglow of sGRB 050709 measured by VLT on July 12, 2005, the data are taken from \citet{Jin2016}. The best fit with a power law is $\nu^{-0.04}$, or with a thermal spectra is temperature $T=7339 K$. Both fits are too poor to be acceptable.}
\end{center}
\end{figure}

\section{An optical flare model for the line-like VLT SED}

In \citet{Jin2016}, the peculiar VLT optical afterglow data has been attributed to the intrinsic spectrum of a single radiation component. In view of that the VLT observations in $I$, $R$ and $V$ bands were carried out in a time interval rather than ``fully simultaneously", in this work we consider an alternative possibility, in which the data may b interpreted assuming an optical flare, with a flux comparable to the ``afterglow" component, appeared after the VLT $I$-band measurement but before the $R$-band measurement on July 12th 2015. It is worth noting that distinct optical flares in late afterglow of GRBs had been well detected \citep[e.g.,][]{Greiner2009}, which had been interpreted as the central engine re-activity \citep[e.g.,][]{Gao2009}. In such a scenario, the peculiar VLT optical SED is not intrinsic but an observation bias, the idea is explained in detail as the following. The time interval of the VLT observations on July 12 2005 is about half an hour, which is much shorter than the measurement time $t\sim 2.5$ days. Hence in such a short time interval the forward shock afterglow emission can be taken as stable and the emission in $V/R/I$ bands are denoted as $f_{\rm V,fs},~f_{\rm R,fs},~f_{\rm I,fs}$, respectively. If there was an optical flare taking place after the $I$-band measurement but before the $V$ and $R$ band measurements, the contribution of the flare emission in the recorded $V/R/I$ fluxes are $f_{\rm V,fl},~f_{\rm R,fl},~0$, respectively. The measured fluxes are hence $f_{\rm V,fs}+f_{\rm V,fl},~f_{\rm R,fs}+f_{\rm R,fl},~f_{\rm I,fs}$, respectively. For illustration one can consider the specific scenario of $f_{\rm V,fs}\approx 2f_{\rm V,fl}$ and $f_{\rm R,fs}\approx f_{\rm R,fl}$. For the forward shock emission it is well known that $f_{\rm I,fs}\approx (\nu_{\rm R}/\nu_{\rm I})^{\beta}f_{\rm R,fs}\approx 1.2f_{\rm R,fs}$ and $f_{\rm V,fs}\approx (\nu_{\rm R}/\nu_{\rm V})^{\beta}f_{\rm R,fs}\approx 0.85f_{\rm R,fs}$, where $\nu_{\rm R}$ and $\nu_{\rm I}$ are the $R$ and $I$ band frequencies, respectively and $\beta\sim 1$. The ``recorded" $I/R/V$ flux ratios are $\sim1.2:2:1.3$, significantly different from the intrinsic SED and may account for our observation,  although both the intrinsic spectra of the afterglow and the flare are power law. If by chance $f_{\rm V,fl}=0$ (i.e., the optical flare took place after the observations in $I/V$ bands), the ``recorded" SED will be a more distinct broadened ``line". Though in this work we focus on the interpretation of the line-like VLT SED of GRB 050709, for illustration in Fig.2 we show in various conditions what will happen in the SED measurement based on the photometric data collected separately in a few bands. Clearly the observed SED could be very peculiar in comparison with that of the regular afterglow spectrum.

\begin{figure*}[ht!]
\label{fig:FLARE}
\begin{center}
\includegraphics[width=0.9\textwidth]{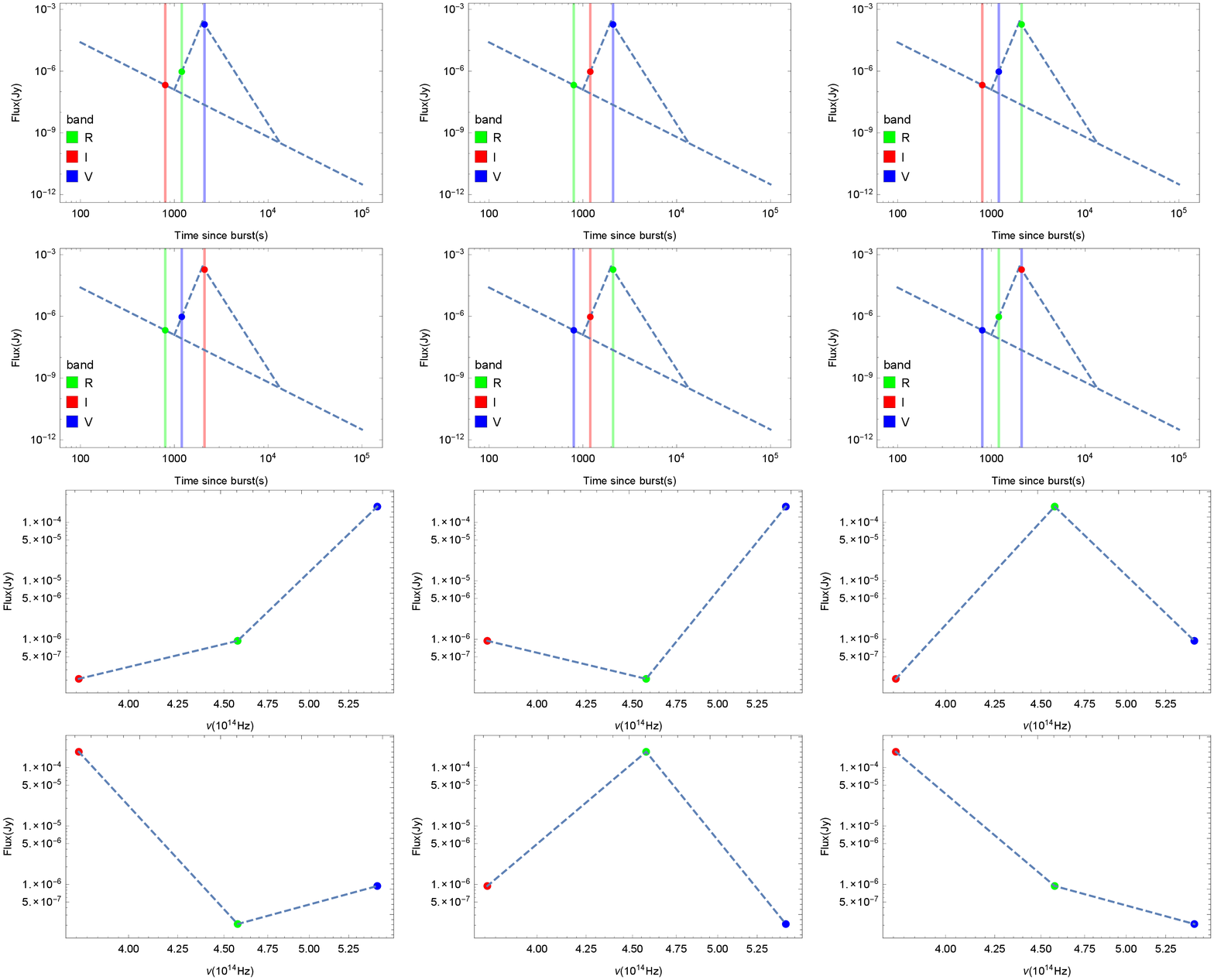}
\caption{A diagrammatic sketch to show possible peculiar SEDs yielded in the superposition of an optical flare with the forward shock afterglow emission.   Here $a', b', c', d', e'$ and $f'$ are the SEDs corresponding to the light curves $a,~b,~c,~d,~e$ and $f$, respectively.}
\end{center}
\end{figure*}

\begin{figure}
\label{fig:OBS}
\begin{center}
\includegraphics[clip, width=0.45\textwidth]{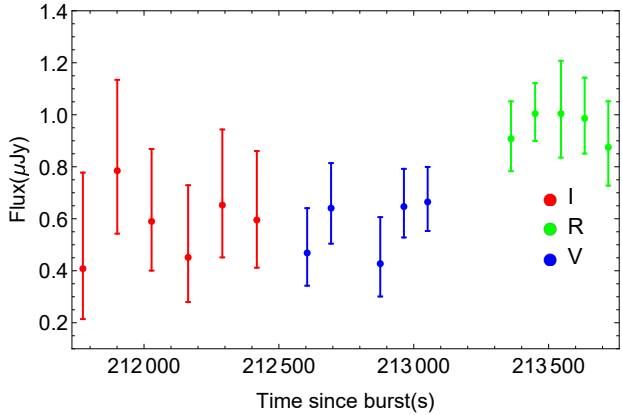}
\caption{The optical observations of GRB 050709 on July 12 2005. Note that each exposure has been analyzed individually.}
\end{center}
\end{figure}

In the current case, the most direct evidence for presence of an optical flare could be the identification of significant flux variability in $R$ and $V$ bands. For such a purpose, we have analyzed the VLT $I/R/V$ band data of GRB 050709 measured on July 12, 2005. The analysis is same as \citet{Jin2016} except in this work we analyze each individual exposure before combine them into a single image in each band. The results are plotted in Fig.\ref{fig:OBS}. It is worth noting that one $V$-band and one $R$-band exposure have been discarded because of their poor quality. For the exposures in $V$ band, the resulting signal to noise is not high enough to reliably check the possible evolution. The signal to noise of the $R$-band exposures is better however one can claim neither there is distinct flux variability nor the flux is stable.

Below we focus on a more specific scenario and try to fit the observation data. As usual we assume that the afterglow flux drops with time as
 $\propto t^{\alpha_{1}}$ and there was an optical flare appearing at $t_0$ and peaking at $t_{\rm p}$. The increase of the optical flare is assumed to be $\propto (t-t_0)^{\alpha_{2}}$. The decrease of the optical flare is assumed to be $\propto (t-t_0)^{\alpha_{3}}$. It is worth noting that due to the so-called high latitude emission, the optical decline index $\alpha_3$ can not be larger than $2+\beta_{\rm o}$ (i.e., $\alpha_3\leq 2+\beta_{\rm o}$), where $\beta_{\rm o}$ is the spectral index in the optical band \citep{Fenimore1996,Kumar2001}.
 For the current potential optical flare, $\beta_{\rm o}$ is essentially unknown and we simply assume that $\beta_{\rm o} \approx 1$ and $\alpha_{3}\approx 3$.

We therefore fit the $R-$band data with the following empirical functions
\begin{displaymath}
F_{\rm R} = \left\{ \begin{array}{ll}
F_{1} t^{\alpha_{1}}, & \textrm{$0<t<t_{0}$}; \\
F_{1} t^{\alpha_{1}} + F_{2} (t-t_{0})^{\alpha_{2}}, & \textrm{$t_{0}<t<t_{\rm p}$}; \\
F_{1} t^{\alpha_{1}} + F_{3} (t-t_{0})^{-3}, & \textrm{$t>t_{\rm p}$}.
\end{array} \right.
\end{displaymath}
To include the contribution of the VLT $I/V$ band data, we convert such emission into $R-$band with an optical spectrum $F_\nu \propto \nu^{\beta_{1}}$ (here, $\beta_{1}=(\alpha_{1}+1)/2$).
The fit to these data yields the physical parameters reported in Table.1, and the results are shown in the insert of Fig.4. These parameters are typical in comparison with other 
 optical flares in GRB afterglows \citep{Li2012}.

\begin{table}
\label{tab:PAR}
\caption{The fit parameters of the afterglow and the flare components.}
\begin{center}
\begin{tabular}{cccccc}
\hline
\hline
Parameter & $\alpha_{1}$ & $\alpha_{2}$ & $\beta_{1}$  & $t_{\rm p}$  & $t_{0}$\\
\hline
Value & $-2.38$ & $2.52$ & $-0.689$ & $213550.8$s & $212683$s\\
\hline
\end{tabular}
\end{center}
\end{table}

\begin{figure}[ht!]
\label{fig:LC}
\begin{center}
\includegraphics[width=0.45\textwidth]{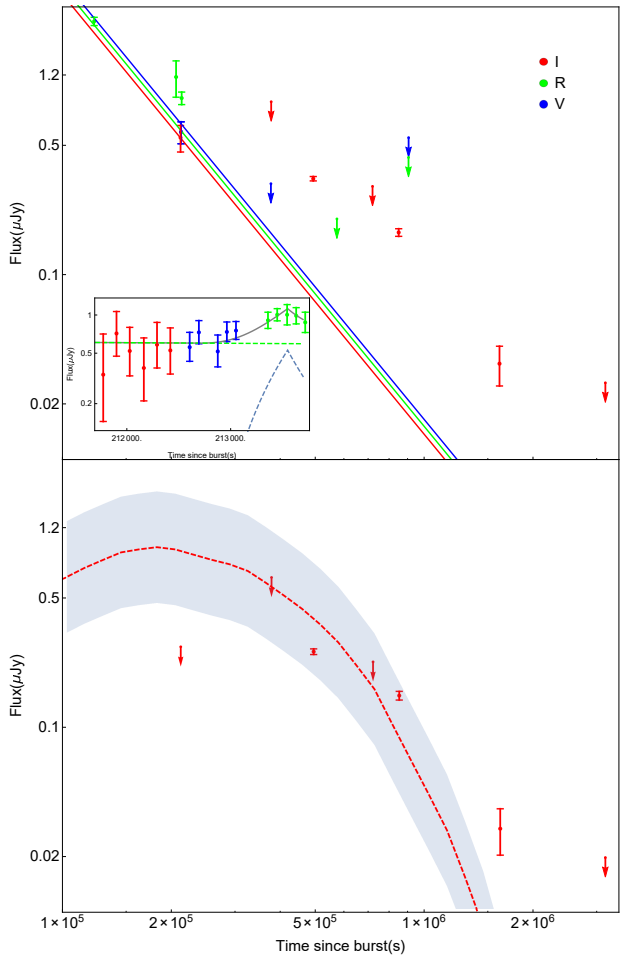}
\caption{Fit to the GRB 050709 afterglow lightcurve with an optical flare. In the upper panel, the solid lines are the ``suggested" lightcurves of the afterglow. In the insert, the VLT data are totally corrected into $R$-band, the green and blue dashed lines and the grey solid line represent the afterglow, optical flare and the total composition respectively. In the lower panel, after the afterglow and the flare components have been subtracted, later HST F814W-band lightcurve is still in significant excess. An macronova model lightcurve (the dashed line for the flux prediction and the shadow region for the model uncertainty) adopted in \citet{Jin2016} is also shown for illustration.}
\end{center}
\end{figure}

\section{Is the macronova emission still needed?}
So far we have shown that the line-like SED measured by VLT on July 12 2005 may be attributed to the superposition of an afterglow component with an optical flare. One would then ask whether a macronova is still needed or not in such a scenario. For such a purpose we subtract the $I-$band/F814W-band ``afterglow" emission,
based on the $R-$band decay $F_{\rm R}\propto t^{-2.38}$ and an optical spectrum $f_{\nu}\propto \nu^{-0.69}$.
Interestingly we have found out that the HST F814W-band data is significantly in excess of such an ``afterglow" component, favoring the presence of a macronova in a wide time interval (from $t\sim 5$ days to $19$ days after the GRB trigger), see Fig.4.

\section{Discussion}

The flares are believed to be produced by late activity of the GRB central engine~ \citep[e.g.,][]{Fan2005,Zhang2006}, i.e., the GRB central engine restarted possibly due to the fallback accretion. One is obliged to accept a more complicated afterglow picture, namely, that the observed afterglow emission is a superposition of the traditional external shock afterglow and an afterglow related to the late central engine activity~ \citep[e.g.,][]{Zhang2006}. If an optical flare opportunely occur during the observations,  a very unusual SED is measured. We discuss six different SEDs when an optical flare is observed at different time using three different bands,  results are compatible with a line-like structure, or an absorption line or even a rising power law (see Fig.2 for illustration).

The three VLT $I/R/V$-band observations of GRB 050709 afterglow on July 12, 2005 constitutes a peculiar SED, which has been interpreted as a broad-line-like spectrum from an iron-group-element-dominated macronova \citep{Jin2016}, which could arise from an accretion disk wind \citep{Metzger2009} in which the heavier r-process elements are depleted because strong neutrino irradiation from a remnant neutron star or the accretion torus can increase the electron fraction of the disk material.
Alternatively as shown in this work that the SED could be due to an flare during the observation. It is hard to distinguishing between these two possibilities due to the limitation of the observations, but both models favor a macronova component that was dominant in the time range between 6 to 19 days. Future detection of a similar SED or another type of SED built in Section 3 may strengthen one of these possibilities. Also, having better timing and wavelength coverage can help. For example, split one round filter shift into more, or make multi-band observation simultaneously. However, to be more reliable, we suggest to split one-round filter shift into two-round, or better if a multi-band (for example $I/R/V$) camera which can make the observations simultaneously on an 8-m telescope. In the future, for similar sources ($R\sim24$mag), spectra can be directly derived from the ELT-class (European Extremely Large Telescope) telescopes, which may finally solve this puzzle. An interesting possibility is that the sub-relativistic neutron-rich ejecta from the compact object mergers may have heavy or lighter composition in different directions and the resulting signal may be a combination of macronova resulting from those  \cite[e.g.,][]{Metzger2014,Kasen2015}. Spectroscopic information for  these faint signals could allow a better understanding of the phenomena, and be useful for understanding the formation of heavy elements via neutron star binary merger.

\section*{Acknowledgements}
We thank Dr. Yi-Zhong Fan for the stimulating discussion and for polishing the presentation. This work was supported in part by the National Basic Research Programme of China (No. 2013CB837000 and No. 2014CB845800), NSFC under grants of 11361140349 (Joint NSFC-ISF Research Program funded by the National Natural Science Foundation of China and the Israel Science Foundation), 11273063, 11433009 and 11103084, the Chinese Academy of Sciences via the Strategic Priority Research Program (Grant No. XDB09000000) and the External Cooperation Program of BIC (No. 114332KYSB20160007).

\end{document}